# Complementarity between private and public investment in R&D: A Dynamic Panel Data analysis

Tarek Sadraoui* & Naceur Ben Zina

*Abstract*— This paper investigates the relationship between private and public investment in R&D, while taking into account the effect of several instruments policies such as subsidies and taxes. We design a new look of knowledge spillovers and R&D cooperation to explain the contribution of public and private R&D on growth. We propose a heterogeneous dynamic panel data model to consider the endogenous effect of R&D investment. We also distinguish between the estimated long and short run results. Our results based on a sample of 23 countries over the period 1992-2004 indicate that both public and private investments in R&D are complementary. By establishing an endogenous growth model, the estimates indicate that public and private R&D depends on the host country's human capital investment. Results indicate that foreign direct investment is a more significant spillover channel than imports.

## I. Introduction

Is public R&D complementary to private R&D, or does it substitute for and tend to "crowd out" private R&D? Conflicting answers are given to this question. We survey the body of available econometric evidence accumulated over the past 16 years. A framework for analysis of the problem is developed to help organize and summarize the findings of econometric studies based on dynamic panel data from various countries[1]. We conclude by offering suggestions for improving future empirical research on this issue.

Most people think that government R&D activities contribute to innovation and productivity, many economists and policymakers have grown frustrated with the paucity of systematic statistical evidence documenting a direct contribution from public R&D (see Paul et al,. [24]. Econometric findings concerning the productivity growth effects of R&D seems to be that there is a significantly positive and relatively high rate of return to R&D investments at both the private and social levels (Cassiman and Veugelers, [8]). In a recent survey, Paul et al,. [24] suggest that the especially pronounced differential over the returns on tangible capital investments observed at the private level may reflect individual firms' perceptions of especially high private risk in the case of R&D. Public funding of R&D can contribute indirectly, by complementing and hence stimulating private R&D expenditures.

Our approach will be to adopt a new econometric approach using a dynamic panel data studies to analyse if public investment in R&D are complement or substitute for private investment in R&D. In literature review, we can conclude that the majority of the econometric studies are concentrated on the impact of public R&D contracts and grants upon private R&D investment by manufacturing firms and industries (see for example, Lach [20], Christopher [10] and Eric [12]).

The object of our paper is to give the theoretical and empirical arguments which allow a satisfactory apprehension of the role that the authorities must play in the fields of research and innovation. The activity of R&D represents a significant source of development of new knowledge and technological innovation (Guellec and Van-Pottelsborghe, [14]). The effort towards activity of R&D involves with a great importance and this through several resources devoted to the various sectors and institutions of research. Expenditure of research and development especially constitute a principal source of growth of productivity for innovating countries. Whereas, Sigrid [23] and Ting [31] suggest that, for countries, where the activity of R&D misses almost technological knowledge and innovations of which they profit are generally resulting from the importation of equipment and goods of intensives investments in technical progress. At the same times, Chaturvedi and Chataway [9] recommend that knowledge capabilities and knowledge management can be considered as key resources for firms in both developed and developing countries.

## II. R&D investment and the government's R&D policies

Today, we can observe an expansion of policies of innovations in the developed countries which devote great investment for R&D. What proves the creation of the climates favourable to the level of these countries for the innovation? It is significant that during these last years, companies of high technology or advanced technology's (pharmaceutical, aeronautical…) expenditure of research and development increased significantly. The role of the governmental policies as regards R&D is not to neglect. Indeed, the policies of innovation define specific actions of the State, which must encourage the accumulation of a qualified labour on the one hand, and to help the companies to prospect the markets on the other hand. This justifies the need for the public administrations for supporting the R&D.

Thus, which are the reasons of the government aid and the mechanisms the alternate ones available to the public administrations to support the R&D? To answer these questions we try to analyze the justification of the government aid with the R&D starting from the economic theories of growth.

### 2.1 Neoclassic growth theory

For Neo-classic theory of growth, technical progress is supposed to be exogenous factors. With the balance of long term, population growth and technical progress determine the level of the growth rate. This implies, according to the basic assumptions, that the long-term growth rate is stable, and given in an exogenous way. Within this framework, the impact of an action of the authorities is practically ignored.

The neoclassic theory of the growth supposes that the economy starts from a weak relationship between capital and labour. Just as the marginal returns on capital are decreasing. What reduces the encouragement to be invested in the new capital? Thus each new unit of capital produces a lower income and less large savings. In the long run, there will be absence of incentive to invest. In short, we can say that the assumptions which underlie the neo-classic theory are not realistic. The technological change is not always an exogenous factor outside the market, determined by an unknown process. To the 20th centuries, a good number of discoveries and progress were carried out in the commercial sector by companies with lucrative goal and not by public administrations or universities where research is directed by non-commercial forces. Markets are seldom in perfect competition, moreover, the private sector is not capable to produce all the desired goods and services, because some of them are goods public and certain others produce external effects.

*2.2 Endogenous growth theory*

The endogenous theory of growth recommends the relaxation of certain neo-classic assumptions and incorporates the failures of the market. However, the economic growth in the long run is directed by the accumulation of the factors of production founded this faith on knowledge, in particular, human capital, training, R&D and innovation (Griliches, [13]). The endogenous models of growth are characterized by a great diversity of the resources selected: The investment in physical capital, in human capital, public capital, and labour division, learning by doing, research and the technological innovation (Romer, [27]).

2.3 R&D investment and market imperfection

Economic theory and empirical proof show that technical progress, because of its incidence on the factors of production, constitutes key element in the long run determining economic growth; in certain countries, it represents even the most significant element. However, it is not a question of an economic justification of the official intervention for allocate the resources in favour of R&D. But, this intervention in a market economy is justified by incapacity of market to distribute resources in an efficient or acceptable way as regards social aspects. With regard to the investment in R&D, external effects and market imperfections testify the incapacity of market, and the effects are felt not only beyond particular companies but also beyond national borders.

In a market economy, a company will not invest in a project if it knows that it can not adapt the possible receipts, however if it cannot adapt a portion of these receipts, it will invest if this portion is enough to make a profitable investment. Asymmetrical information and imperfect competition constitute two other kinds of imperfections of market involving under investment in R&D. For example, asymmetrical information prevents effective operation of capital market. Indeed, it can involve rationing of appropriations as well as abandonment of investments in R&D in projects with strong chances of success thanks to the plan of financing, and the continuation of investments in the project having weak chances of success.

III. RESEARCH METHODOLOGY

Our study contributes to the empirical literature -which is discussed here- on the analysis of the existence of a relation between private and public investment in R&D and their real effect on economic growth; do public funds substitute or complement private R&D expenditure?

We derive our econometric specification from a function including interactions between internal and external R&D in the augmentation of the knowledge stock. The model also takes into account potential productivity convergence by including lagged productivity levels. Our study's inferences are based on a dynamic panel data model, which allows us to control for the existence of unobserved fixed effects that are likely to affect R&D decisions. Estimation is carried out by several consistent dynamic panel data methods, among which generalized method of moments, which allows for the presence of weakly endogenous explanatory variables. In this way the analysis can take into account both degree and possibility effects of R&D to address the issue of optimal combinations of R&D expenditures.

In this paper, we contribute the first panel data study exploring complementarity between public R&D and private R&D in a dynamic panel framework. We examine the impact of internal and external R&D on economic growth in sixteen-year panel for 23 developed and developing countries.

IV. COMPLEMENTARITY VERSUS SUBSTITUTABILITY BETWEEN PRIVATE AND PUBLIC R&D

Theoretical work did not succeed in slicing on favourable or unfavourable effect using certain political instruments on the level of R&D in private sectors. The results of each model strongly depend on its structure and its assumptions. Empirical work, leads to homogeneous results and identifies a positive effect of public R&D on that private (Paul, et al,. [24]). With an aim of knowing the relation between public and private R&D we give an overall picture of the activities of R&D in world. Indeed, in this section, we attach more importance to activity of public and private R&D in the most significant poles in world.

Through time and with the improved scientific methods in particularly studies of Jason, et al., [19], it became clearly that the final situation towards the effect of the public funds of R&D cannot be made. Thus, in general, two fields can be identified and which are used to analyze the relation between private investment and public in research and development with knowing quantitative and qualitative studies: On the one hand; for the qualitative studies, data are frequently based on the investigations. On the other hand; for the quantitative studies, they are based on macro and micro-economic information of a significant number of companies (Cassiman and Veugelers, [8]).

Today, several activities of R&D are carried out on the level of the services sector. On the one hand, this is due to the external sources of the strategies of manufacturing industries in the Eighties. On the other hand, the transformation of information and technology of communication get more opportunities for innovating sectors. So the governments help more and more activities of R&D in several sectors with an aim of stimulating technological performances of their countries. Thus, several examples can be quoted. At this level, for the Nineties and more precisely in 1999, the total expenditure of R&D of Germany is 47 billion dollars where 66% of this amount is invested by private industries, 18% by government and the remainder are invested by foreign companies.

Thus, Claudia, [11] suggest that an international comparison on behalf of public programs of R&D shows that Germany is one of principal countries which grant funds for the technological performance. At this level, manufacturing industry plays a very significant role concerning R&D. For example, the strategic planning of the national research evaluation in Thailand as indicates it Jarunee, [18] is to allocate the budget to support the research programmes and projects. Jarunee suggest that to improve the model evaluation framework for R&D investments, the public hearing forum was organised. From there, a question emerges up to what point evolution of public funds of R&D makes it possible to stimulate R&D carried out by private sector, and on which level results are checked? Recently, an econometric micro study tackled the question of the impact of political instruments about activity of R&D deprived on the level of companies.

Several other studies are more precisely interested in testing the effects of public subsidies in R&D on the amount of deprived investment see for example Lai et al., [21], Hans and Almas, [16] and Christopher, [10]. The major goal of these studies is to know if public subsidies of research and development can have an effect of reduction or increase in the expenditure of R&D. Most of results suggest that public subsidies of R&D on the level of several industries showed that there is a small tendency to the effect of ousting "Crowding out". In addition, it seems not to have any effect or degree of Complementarity.

In the following section we empirically test fundamental relation which we seek to analyze in the case of 23 countries for the period 1992-2004, in other words we test the existence of a relation of Complementarity and to check this result.

V. EMPIRICAL VALIDATION: DYNAMIC PANEL DATA

The objective of this work is to test the impact of an action of public policies empirically on the evolution of R&D in private sector while trying to surmount limits. The modelling that we follow to measure the effect of the R&D deprived on the public one; while taking into account some determinants of private R&D; is the one of Bettina and al,. (2002). This modelling has also been applied by: Busom, [7] and Lach, [20]. The gait of these authors can be summarized like follows:

**Private R&D = ß * public R&D + control variables + e**

The underlying logic is simple: If the coefficient β* has a positive sign we can say that public R&D are complementary for private R&D. In other words, an increase of 1% of public research and development level entails a growth of β*% of private R&D. On contrary, if β* has a negative sign we can say that there is a relation of substitutability between public and private R&D. In this part we try, to give a general setting for the models to estimate while putting accent on some remarks and inconveniences of these models. We apply a dynamic panel data model. Finally, after having estimated the model we analyze results.

5.1 Dynamic panel data: Definition and evaluations method

Dynamic models are characterized by presence of one or several endogenous variables delayed among explanatory variables. Our specified model is a dynamic panel model is given by:

$$y_{it} = \alpha y_{it-1} + \beta' x_{it} + v_{it} \quad (1)$$

Under another forms one was writing our model as below:

$$R_{it} = \alpha R_{it-1} + \beta_1 G_{it} + \beta_2 M_{it} + \beta_3 VA_{it} + \beta_4 IDE_{it} + \mu_i + v_{it} \quad (2)$$

Where;

$y_{it-1}$ : Endogenous variable appears in the regression as being a retarded explanatory variable. In other words, present stocks of research and development of country (i) are explained by stocks of research of the period (t-1).

X: Represent the vector of exogenous variables; these variables are added value (VA), public research (G), import (M), foreign direct investment (IDE) and private research;

(α, β): Designate parameters to estimate;

$\mu_i$ : Constitute individual heterogeneity as: $\mu_i$ i.i.d. ~ N [(0, 1)];

And: $v_{i,t}$ is stochastic term as: $v_{i,t}$ ~ i.i.d. [(0, 1)].

$y_{i,t}$ is the logarithm of volume of R&D in country (i).

$x_{i,t}$ is determinant vector of R&D.

$\mu_i$ is the specific effect of country (i). This specific effect can be a stationary or uncertain effect.

*5.2 Evaluation Method*

The evaluation of the model by traditional methods (Ordinary Least Square "OLS" and within) gives biased and non convergent values because of inter-relationship between retarded endogenous variable and individual heterogeneity. We try to demonstrate for the case of a simple model the inconveniences of these methods of evaluations.

For dynamic panel model, Within transformations and Ordinary Least Squares are biased and non-convergent estimators.

We assume the simple specification:

$$y_{it} = \alpha y_{it-1} + \mu_i + v_{it}$$

Let us pose $|\alpha| \prec 1$ and that $y_{it}$ is stationary then:

$$\hat{\alpha} = \frac{\sum_{i=1}^{N} \sum_{t=1}^{T} y_{it} y_{it-1}}{\sum_{i=1}^{N} \sum_{t=1}^{T} y_{it-1}^2} = \alpha + \frac{\sum_{i=1}^{N} \sum_{t=1}^{T} (\mu_i + v_{it}) y_{it-1}}{\sum_{i=1}^{N} \sum_{t=1}^{T} y_{it-1}^2}$$

$\hat{\alpha}$ is convergent if

$$p\lim_{N \longrightarrow \infty} \frac{1}{NT} \sum_{i=1}^{N} \sum_{t=1}^{T} (\mu_i + v_{it}) y_{it-1} = 0$$

$$p\lim_{N \longrightarrow \infty} \frac{1}{NT} \sum_{i=1}^{N} \sum_{t=1}^{T} (\mu_i + v_{it}) y_{it-1}$$

$$= \frac{1}{T} \frac{1-\alpha^{2T}}{1-\alpha} \text{cov}(y_{i0}, \mu_i) + \frac{\sigma_\mu}{T(1-\alpha)^2} [(T-1) - T\alpha + \alpha^T]$$

$$= p\lim_{N \longrightarrow \infty} \frac{1}{NT} \sum_{i=1}^{N} \sum_{t=1}^{T} y_{it-1}^2$$

$$\frac{1}{T} \frac{1-\alpha^{2t}}{1-\alpha} \frac{\sum_{i=1}^{N} y_{i0}}{N} + \frac{\sigma^2 \mu}{T(1-\alpha)^2} \frac{1}{T} \left[ T - 2\frac{1-\alpha^{2t}}{1-\alpha} + \frac{1-\alpha^{2t}}{1-\alpha^2} \right]$$

$$+ \frac{2}{T(1-\alpha)} \left[ \frac{1-\alpha^{2t}}{1-\alpha} - \frac{1-\alpha^{2t}}{1-\alpha^2} \right] \text{cov}(y_{i0}, \mu_i) +$$

$$+ \frac{\sigma^2 \mu}{T((1-\alpha)^2)^2} \left[ (T-1) - T y^2 + \alpha^{2T} \right]$$

In summary, the bias is positive and increases with the variance of the specific effect. Indeed, $y_{i,t}$ is function of $v_{i,t}$ and $y_{i,t-1}$ is also. $y_{i,t-1}$ is an explanatory variable correlated with stochastic term. It introduces a bias in the value of ordinary least squares. Even as putting hypothesis that stochastic terms are not correlated, this value is non-convergent.

For within case we consider the following transformation:

$$y_{it} - \bar{y}_{i.} = \alpha(y_{it-1} - \bar{y}_{i.-1}) + (v_{it} - \bar{v}_{i.-1})$$

With:
$$\begin{cases} E(v_{it} v_{js}) = 0 & \text{si } i = j \text{ et si } t = s \\ E(v_{it} v_{js}) \neq 0 & \text{if not} \end{cases}$$

While posing as in Baltagi [3]:

$$\bar{y}_{i.-1} = \frac{\sum y_{it-1}}{T-1}$$

Thus we can write that:

$$\hat{\alpha} = \frac{\sum_{i=1}^{N} \sum_{t=1}^{T}(y_{it} - \bar{y}_{i.})(\bar{y}_{it-1} - \bar{y}_{i.-1})}{\sum_{t=1}^{T} \sum_{i=1}^{N}(\bar{y}_{it-1} - \bar{y}_{i.-1})^2}$$

$$= \alpha + \frac{\sum_{i=1}^{N} \sum_{t=1}^{T}(\bar{y}_{it-1} - \bar{y}_{i.-1})(v_{it} - \bar{v}_{i.})/NT}{\sum_{i=1}^{N} \sum_{t=1}^{T}(\bar{y}_{it-1} - \bar{y}_{i.-1})^2/NT}$$

The numerator is convergent when the second term converges towards zero.

The numerator of the second term:

$$p\lim_{N \to \infty} \frac{1}{NT} \sum_{i=1}^{N} \sum_{t=1}^{T} (\bar{y}_{it-1} - \bar{y}_{i.-1})(v_{it} - \bar{v}_{i.})$$

$$= \frac{-s_v^2}{T^2} \frac{(T-1) - Ta + a^T}{(1-a)^2}$$

And the denominator

$$p\lim_{N \to \infty} \frac{1}{NT} \sum_{i=1}^{N} \sum_{t=1}^{T} (\bar{y}_{it-1} - \bar{y}_{i.-1})^2$$

$$= \frac{-s_v^2}{1-a^2}\left\{1 - \frac{1}{T} \frac{2a}{(1-a)^2} \frac{(T-1) - Ta + a^T}{T^2}\right\}$$

The estimator of Least Square Dummy variable is convergent if T is infinite. If T is fixed, N, the estimator is non-convergent. Our model should not be estimated by the method of OLS and LSDV due to the fact that estimating by these methods led to ad hoc results. Which are then adequate methods to estimate our model? We propose below two methods which consist in obtaining consistent estimators.

### 5.3 Anderson and Hsiao Method

Anderson and Hsiao [1] proposed, initially, to write the model from first difference to eliminate individual heterogeneity. They propose for the transformation two instruments.

$$\hat{\alpha}_{vi}(1) = \frac{\sum_{i=1}^{N} \sum_{t=3}^{T}(y_{it} - y_{it-1})(y_{it-2} - y_{i.-3})}{\sum_{i=1}^{N} \sum_{t=1}^{T}(y_{it} - y_{i.-2})(y_{it-2} - y_{i.-3})} \quad (3)$$

And

$$\hat{\alpha}_{vi}(2) = \frac{\sum_{i=1}^{N} \sum_{t=3}^{T}(y_{it} - y_{it-1}) y_{it-2}}{\sum_{i=1}^{N} \sum_{t=1}^{T}(y_{it} - y_{i.-2}) y_{it-2}} \quad (4)$$

The two values are convergent when N and $T \to \infty$. However, an inter-relationship always persists between endogenous variable in first difference and residual term. Authors proposed to resort to the method of instrumental variables to surmount this problem. Thus, they propose to use instrument endogenous variable with two lags or his first differences. These instruments are correlated with explanatory variable and are not with residual term. To get more efficient results, Arellano and Bond [2] approach permits to get a value of generalized moments "GMM" more efficient.

### 5.4 Arellano and Bond Approach

Arellano and Bond [2] are the first in 1991 that proposed an extension of GMM introduced initially by Hansen [15], to the case of panel data for a simple model AR (1):

$$y_{it} = \alpha y_{it-1} + \mu_i + v_{it} \quad (5)$$

Where $|\gamma| \prec 0$

We consider the case where temporal dimension is small while individual dimension (N) is important. We consider that individual effects are stationary and we assume traditional hypotheses of residues:

In difference models (5) can be written as below:

$$\Delta y_{it} = \alpha \Delta y_{it-1} + u_{it} \quad |\gamma| \prec 0 \quad (6)$$

Where $u_{it} = v_{it} - v_{it-1}$.

We test for every individual of the linear restrictions of type:

$$E\left[(\Delta y_{it} - \alpha \Delta y_{it-1}) y_{it-j}\right] = 0 \quad \text{for } j = 2,...,t; t = 3,...T \quad (7)$$

The gait of Arellano and Bond, in presence of the exogenous variables, consists in estimating the model in difference:

$$\Delta y_{it} = \sum_{k=1}^{p} \alpha_k \Delta y_{i(t-k)} + \beta'(L) X_{it} + \Delta v_{it} \quad (8)$$

Moment conditions and instruments matrix are given respectively by:

$$\begin{cases} E(y_{it-\tau} \Delta v_{is}) = 0 & \text{pour } \tau \geq 2, t = 2,3,...T \\ E(X_{it-\tau} \Delta v_{it}) \neq 0 & \text{pour } \tau \geq 2, t = 2,3,...T \end{cases} \quad (9)$$

$$Z = \begin{pmatrix} Z_{ip} & 0 & 0 & \cdots & 0 \\ 0 & Z_{ip+1} & \cdots & & \cdot \\ \cdot & \cdot & \cdot & \cdot & \cdot \\ \Delta X_{ip+2} & \Delta X_{ip+3} & \cdots & & \Delta X_{iT} \end{pmatrix} \quad (10)$$

The preceding dynamic model (8) can be rewritten for each individual in the following form:

$$y_i = W_i \delta + \tau_i \mu_i + V_i \quad (11)$$

Where $\tau$ is a vector of parameter and $W_i$ is a matrix that contains the retarded dependent variable and explanatory variables. The method proposed by these author's permits to get a GMM in two stages is written in following form:

$$\hat{\delta} = \left[\left(\sum_i W_i^{*'} Z_i\right) A_N \left(\sum_i Z_i' W_i^*\right)\right]^{-1} \left(\sum_i W_i^{*'} Z_i\right) A_N \left(\sum_i Z_i y_i^*\right) \quad (12)$$

However, to have previous value GMM, it is necessary to pass by a first stage that consists in making wished transformation (first difference or orthogonal deviation), to find and to use instruments matrix and to achieve a first evaluation named "evaluation of first stage". This stage corresponds to an evaluation that permits to provide estimated residues after transformation. In the first stage, the values are gotten while using $H_i$ as:

$$H_i = \hat{v}_i^* \hat{v}_i^{*'} \quad (13)$$

$$H_i = \begin{pmatrix} 2 & -1 & \cdots & & 0 \\ -1 & 2 & \cdot & & \cdot \\ 0 & \cdot & \cdot & & -1 \\ 0 & 0 & \cdots & -1 & 2 \end{pmatrix}$$

And

$$A_N = \left(\frac{1}{N} \sum_i Z_i' H_i Z_i\right)^{-1} = Z'(I_N \ddot{A} H) Z \quad (14)$$

The objective of transformation is, as at Anderson and Hsiao [1], to eliminate individual heterogeneity of the model. The number of instrument increases in the time for every individual. In the case where exist explanatory variables $x_{it}$ in the model correlated with heterogeneity individual $\mu_i$. Optimal instruments matrix corresponding $Z_i$ is equal to:

$$\begin{pmatrix} y_{i1} & x_{i1} & x_{i2} & 0 & 0 & 0 & 0 & \cdots & 0 & 0 & 0 & \cdots & 0 & 0 \\ 0 & 0 & 0 & y_{i1} & x_{i1} & x_{i2} & x_{i3} & \cdots & 0 & 0 & 0 & \cdots & 0 & \cdot \\ \cdot & \cdot & \cdot & \cdot & \cdot & \cdot & \cdot & & \cdot & \cdot & \cdot & \cdots & \cdot & \cdot \\ 0 & 0 & 0 & 0 & 0 & 0 & 0 & \cdots & y_{i1} & y_{i(T-2)} & x_{i1} & \cdots & \cdots & x_{i(T-1)} \end{pmatrix}$$

Arellano and Bond [2] propose a test verifying the absence of autocorrelation of first and second order. Thus, if distribution is non auto-correlated, this test gives a value of residues differentiated negative and significant to first order and non significant to the second order. This test that is based on auto-covariance of residues follows a normal law N (0,1) under hypothesis $H_0$. Otherwise, authors propose the test of validity of instruments of Sargan (1988a). The statistical test equal to:

$$S = \left(\sum \hat{v}_i^{*'} Z_i\right) A_N \left(\sum Z_i' \hat{v}_i^*\right) \quad (15)$$

## VI. EMPIRICAL RESULTS

The unit root tests became a current step for analysis of time series stationnarity. However, practical application of these tests on panel data is recent. The tests most frequently used are those of Levin and Lin, [22] and of Im, Pesaran and Shin [17][3]. Recently, several procedures of unit root tests and Cointegration were developed for panel data models. The addition of individual dimension to temporal dimension offers an advantage, in practical application of unit root and Cointegration tests (Pedroni, [25, 26]).

The checking of non-stationary properties for all panel variables leads us to study the existence of a long run relation between these variables. The Cointegration study by applying Pedroni Cointegration tests based on unit root tests on residues estimated. Cointegration tests on panel data consist in testing the presence of unit root in the estimated residues. However, the problem of fallacious regressions, of the time series, also arises in the case of panel data. First step is to test unit root for each of series.

### 6.1 Unit Root Tests
Levin and Lin [22], consider the following model:
$$y_{i,t} = \rho_i y_{i,t-1} + Z'_{it} \gamma + u_{i,t} \quad (i=1, \ldots, N; t=1, \ldots, T) \quad (16)$$

Where, $Z_{i,t}$ is the deterministic component and $u_{i,t}$ is a stationary process. $\mu_i$ is the fixed effect,

The Levin and Lin, [22] tests assume that $u_{i,t}$ are iid $(0, \sigma_u^2)$ and $\rho i = \rho$ for all i. The LL test is restrictive in the sense that it requires $\rho$ to be homogeneous across i. Im, Pesaran and Shin [17] (IPS) allow for a heterogeneous coefficient of $y_{i,t-1}$ and propose an alternative testing procedure based on averaging individual unit root test statistics. IPS suggested an average of the augmented Dickey-Fuller (ADF) tests when $u_{i,t}$ is serially correlated with different series. Correlation properties across cross-sectional units, i.e;

$$u_{i,t} = \sum_{j=1}^{p_i} \alpha_{ij} u_{it-j} + \varepsilon_{it}.$$

Substituting this $u_{i,t}$ in (1) we get:

$$y_{i,t} = \rho_i y_{it-1} + \sum_{j=1}^{p_i} \alpha_{ij} \Delta y_{it-j} + z'_{it} \gamma + \varepsilon_{it} \quad (17)$$

The null and for all countries i the alternative hypothesis are:
Ho: $\rho_i = 1$
Ha: $\rho_i < 1$
For at least one i, the IPS t-bar statistic is defined as the average of the individual ADF statistic as:

$$\bar{t} = \frac{1}{N} \sum_{i=1}^{N} t_{\rho i}$$

Where $t\rho i$ is the individual t-statistic of testing Ho: $\rho i = 1$ in (18). It is known for a fixed N as $T \to \infty$

$$t_{\rho i} \Rightarrow \frac{\int_0^1 W_{iz} dW_{iz}}{\left[\int_0^1 W_{iz}^2\right]^{1/2}} = t_{iT} \quad (18)$$

IPS assumes that $t_{iT}$ are iid[4] are have finite mean variance. Then;

$$\frac{\sqrt{N}\left(\frac{1}{N}\sum_{i=1}^{N} t_{iT} - E[t_{iT}/\rho_i=1]\right)}{\sqrt{Var[t_{iT}/\rho_i=1]}} \Rightarrow N(0,1) \quad (19)$$

As $N \to \infty$ central limit theorem. Hence

$$t_{IPS} = \frac{\sqrt{N}\left(\bar{t} - E[t_{iT}/\rho_i=1]\right)}{\sqrt{Var[t_{iT}/\rho_i=1]}} \Rightarrow N(0,1) \quad (20)$$

As $T \to \infty$ followed by $N \to \infty$ sequentially, the values of E[tiT/$\rho i$=1] and Var[tiT/$\rho i$=1] have been computed by IPS simulations for different values of T and $\rho'$is. As applying test on our complete model our results is summarized in table 1

## VII. CONCLUSION

In our survey, we tried to put accent on private and public investment in R&D, for the case of 23 countries which presents different levels of R&D. We tried to clarify relation that exists between private and public research. This empirical survey wanted to give account, the effects of different determinants on private investment in R&D and to know if public and private investments in R&D are complement or substitute.

Econometric approach consists in the regression of some measures of private R&D on public R&D with some control variables. The evaluation that we presented in our work corresponds to GMM evaluation in first difference and in orthogonal deviation. We prefer to refer to results of this evaluation because it permits to eliminate rigorous way all bias to none observed individual heterogeneity and offer, a better efficiency of results. Empiric evaluations confirm a positive effect of public R&D in different country (positive and meaningful effect in all evaluations). Results of our empiric survey are relative for our sample and they go in the sense of results of ulterior studies, which showed that there is a positive and meaningful relation between private and public investment in R&D.

All results are in favor of a positive relation between private and public R&D which can be assumed by a complementarities between them. In our study we have indicate that all variables are stationary by an application of unit root tests that's can contribute to search Cointegration relation between them and determinate the number of these relation. Another important think, we can give the impact of public R&D to private R&D for each country to specify the nature of relation and how private R&D contributes for public sector. In summary, all countries must stimulate private sector in R&D activities to promote economic growth and integrate a new innovation system which can go with their own economic environment.

Some studies put in value of other factors that can be important as: competition in the market, public politics and cooperation concerning R&D between firms. Cooperation in R&D is a part of the new strategies developed by firms in more global and competitive economic environment. These last factors are not to disregard and can be subject of a future research concerning the relation between public and private investment in R&D.

APPENDIX

**Table 1** Unit root Tests

| Statistics | R | G | M | VA | IDE |
|---|---|---|---|---|---|
| **Levin-Lin ADF stat** | -2.357 | -0.343 | 1.809 | 1.489 | -1.674 |
| **IPS ADF stat** | -2.296 | 1.773 | 2.175 | 1.640 | -1.572 |